\documentclass[useAMS,usenatbib,usegraphicx]{mn2e}
\usepackage{multirow}
\usepackage{color,graphicx}
\usepackage[letterpaper]{geometry}
\usepackage{longtable,lscape}
\newcommand{\lsim}{\raisebox{-.5ex}{$\,\stackrel{\textstyle <}{\sim}\,$}}
\newcommand{\gsim}{\raisebox{-.5ex}{$\,\stackrel{\textstyle >}{\sim}\,$}}
\newcommand{\mnras}{MNRAS}

\newcommand{\apj}{ApJ}
\newcommand{\apjs}{ApJS}
\newcommand{\apjl}{ApJL}
\newcommand{\aj}{AJ}
\newcommand{\aap}{A\&A}
\newcommand{\aaps}{A\&AS}
\newcommand{\araa}{ARA\&A}
\newcommand{\pasp}{PASP}

\title{The number fraction of discs around brown dwarfs in Orion OB1a and the 25 Orionis group}
\author[J. J. Downes et al.]
{Juan Jos\'e~Downes$^{1,2}$\thanks{E-mail: jdownes@cida.ve, jdownes@astrosen.unam.mx},
Carlos Rom\'an-Z\'u\~niga$^{1}$  \thanks{E-mail: croman@astrosen.unam.mx},
Javier Ballesteros-Paredes$^{3}$ \newauthor
Cecilia~Mateu$^{1,2}$, 
C\'esar~Brice\~no$^{4}$,
Jes\'us~Hern\'andez$^{2}$, 
Monika G.~Petr-Gotzens$^{6}$,\newauthor
Nuria~Calvet$^5$,
Lee~Hartmann$^5$,
and Karina Mauco$^{3}$\\
$^{1}$Instituto de Astronom\'ia, UNAM, Ensenada, C.P. 22860, Baja California,  M\'exico\\
$^{2}$Centro de Investigaciones de Astronom\'{\i}a, AP 264, M\'erida 5101-A, Venezuela\\
$^{3}$Centro de Radioastronom\'ia y Astrof\'isica, UNAM. Apartado Postal 72-3 (Xangari), Morelia, Michoac\'an 58089, M\'exico\\
$^{4}$Cerro Tololo Interamerican Observatory, Casilla 603, La Serena, Chile\\
$^{5}$Department of Astronomy, University of Michigan, 825 Dennison Building, 500 Church Street, Ann Arbor, MI 48109, USA\\
$^{6}$European Southern Observatory, Karl-Schwarzschild-Str. 2, 85748 Garching bei M\"unchen, Germany\\}
\begin{document}
\date{Accepted 2015 ------ --. Received 2015 ------ --; in original form ---- ------ --}
\pagerange{\pageref{firstpage}--\pageref{lastpage}} \pubyear{2015}
\maketitle
\label{firstpage}
\begin{abstract}
We present a study of 15 new brown dwarfs belonging to the $\sim7~\rmn{Myr}$ old 25
Orionis group and Orion OB1a sub-association with spectral types between M6 and M9 
and estimated masses between $\sim0.07~\rmn{M}_\odot$ and $\sim0.01~\rmn{M}_\odot$. By comparing 
them through a Bayesian method with low mass stars ($0.8\lsim \rmn{M}/\rmn{M}_\odot\lsim0.1$) 
from previous works in the 25 Orionis group, we found statistically significant differences 
in the number fraction of classical T Tauri stars, weak T Tauri stars, class II, evolved 
discs and purely photospheric emitters at both sides of the sub-stellar mass limit. 
Particularly we found a fraction of $3.9^{+2.4}_{-1.6}~\rmn{\%}$ low mass stars classified as CTTS 
and class II or evolved discs, against a fraction of $33.3^{+10.8}_{-9.8}~\rmn{\%}$ in the sub-stellar 
mass domain. Our results support the suggested scenario in which the dissipation of discs 
is less efficient for decreasing mass of the central object. 
\end{abstract}
%
%
\section{Introduction}\label{introduction}
%
%
After two decades of studying the properties of young brown dwarfs (BD), there is an agreement that 
these objects and very low mass stars (VLMS) have similar formation processes \citep[e.g.][]{luhman2012} 
although some issues about their early evolution are still matter of intense research.
%
%
For instance, it is well accepted that the number fractions of VLMS and BD harbouring primordial 
discs drop off during the first $\sim10~\rmn{Myr}$ of their evolution \citep[e.g.][]{luhman2012}.
However, there is growing evidence suggesting that the time scale for disc dissipation could 
depend on stellar mass, because the fraction of objects that retain circumstellar discs increases 
for those with lower masses \citep[e.g.][]{luhman2012b}. Such a dependency could have remarkable 
implications on the efficiency for the formation of giant and terrestrial planets around stars of 
different masses \citep[e.g.][]{pascucci2009,pascucci2013}.
%
%
\par
The first evidences of such a dependency came from studies in the stellar mass domain.
%
%
\citet{hernandez2005} found that the inner disc frequency at 
ages between 3 and 10 $~\rmn{Myr}$ in intermediate-mass stars is lower than in low-mass stars
and suggested that it could be a consequence of a more efficient mechanism of primordial 
disc dispersal in the intermediate-mass stars.
%
%
These results were subsequently supported by \citet{megeath2005}, \citet{siciliaaguilar2006}
and \citet{lada2006}. Furthermore, \citet{carpenter2006} studied the number fraction of 
discs surrounding stars in the wide mass range $0.1\lsim \rmn{M}/\rmn{M}_\odot\lsim20$ 
in the Upper Sco association, finding that the fraction of primordial optically thick discs decreases as 
the mass of the star increases. 
%
%
\citet{hernandez2007} studied stars with spectral types K6 to M4 in the Orion OB1 
association and found the maximum of the disc frequency in M0. 
In the younger $\sigma$ Ori cluster \citet{hernandez2007b} reported disc fractions of 
$10~\rmn{\%}$ for $\rmn{M}>2~\rmn{M}_\odot$ and $35~\rmn{\%}$ for $0.1<\rmn{M}/\rmn{M}_\odot<1$. This tendency is also supported 
by results in the $\lambda$ Orionis association from \citet{hernandez2010} who found a disc 
fraction of $\sim6~\rmn{\%}$ for K-type stars and $\sim27~\rmn{\%}$ for M5 stars or later.
%
%
\par
The unprecedented capabilities of the Spitzer Space Telescope \citep{fazio2004,rieke2004} 
and the WISE survey \citep{wright2010} allowed to extend these studies down to the sub-stellar 
mass domain.
%
%
\citet{luhman2008} complemented the results from \citet{hernandez2007b} for the $\sigma$ Ori 
cluster finding that $60~\rmn{\%}$ of the BDs shows IR excesses consistent with discs.
%
%
\citet{riaz2008} studied the TW Hya association and found that $\sim60~\rmn{\%}$ of the BDs show IR excesses, 
against $\sim24~\rmn{\%}$ of the VLMS and \citet{riaz2009} suggest that the longer disc lifetimes in TW Hya 
could be a consequence of its lower BD spatial density. 
%
%
In the same region, \citet{morrow2008} showed that the BDs having irradiated accretion discs do not 
show the silicate emission found in discs around VLMS \citep{uchida2004,furlan2007} and in younger 
VLMS and BDs. They interpreted these results as an indication that grain growth occurs more rapidly 
in discs around BD than in those around stars or that grains grow faster at smaller disc radii as 
suggested by \citet{kessler2007} and \citet{siciliaaguilar2007}.
%
%
\par
Particularly, the $\sim5~\rmn{Myr}$ old Upper Sco association \citep[e.g][]{preibisch2002} has been the subject 
of several studies about the fraction of BD harboring discs. \citet{scholz2007} studied VLMS and BD with 
spectral types from M5 to M9, finding a disc frequency of $\sim37~\rmn{\%}$. They also reported that $\sim30~\rmn{\%}$ 
of such objects also show H$\alpha$ emissions consistent with active accretion.
%
%
\citet{riaz2012} compiled all the spectroscopically confirmed BDs and identified 
discs based on WISE photometry. They found a disc frequency of $\sim28~\rmn{\%}$ and that half 
of the VLMS and BD harboring discs also show signatures of accretion. They did not find any dependence 
of the disc fraction with the stellar number density or the BD/star number ratio and suggested that the 
differences in disc life times could also be a consequence of different BD 
formation mechanisms and/or different initial disc fractions.
%
%
Considering the latest age estimate for that region \citep[$\sim11~\rmn{Myr}$;][]{pecaut2012}, \citet{luhman2012b} 
found that the disc fraction of objects with masses $0.01\lsim \rmn{M}/\rmn{M}_\odot\lsim0.2$ reaches $\sim25~\rmn{\%}$, 
which indicates that such primordial discs could survive for at least $\sim10~\rmn{Myr}$.
%
%
Recently, in this region, \citet{dawson2013} reported that $23~\rmn{\%}$ of the BDs are class II and that $19~\rmn{\%}$ are 
transitional discs.
They also compared their fractions with those for K and M-type stars in Upper Sco, ChaI, IC348 and $\sigma$ Ori 
\citep[][respectively]{luhman2005,damjanov2007,lada2006,hernandez2007} and argued that the correlation 
between the disc fraction and the stellar mass is not statistically significant, and that the average lifetimes 
of discs around such stars could not depend on the stellar mass.
%
%
Using VISTA \citep{emerson2004,emerson2010,petrgotzens2011}, IRAC-Spitzer and WISE photometry \citet{downes2014a} 
studied the $\sim7~\rmn{Myr}$ old 25 Orionis group and Orion OB1a finding a number fraction of candidates low-mass stars
(LMS) showing IR excesses between $\sim7~\rmn{\%}$ and $\sim10~\rmn{\%}$ while for BDs candidates the number fraction increases 
up to $\sim20~\rmn{\%}$ to $\sim50~\rmn{\%}$.
%
%
\par
In summary, there are several indications on a possible dependence of the characteristic time 
scale of disc dissipation with stellar mass, that extends down to the sub-stellar mass regime, 
although some issues related to the estimation of ages and the uncertainties in the number 
fractions need to be clarified. These issues could be addressed by studying different disc 
indicators in slightly more evolved regions ($\rmn{t}\gsim5~\rmn{Myr}$) with consistent age estimations. 
This, together with the use of numerous samples of LMS and BDs and a thorough statistical 
treatment, can provide a robust analysis of the variations of disk fractions with age.

\par
In this paper we present an optical spectroscopic and optical/near-IR photometric study 
of 15 BDs with masses $0.01\lsim \rmn{M}/\rmn{M}_\odot\lsim 0.07$ (spectral 
types between M6 and M9), together with the sample of 77 LMS with masses 
$0.1\lsim \rmn{M}/\rmn{M}_\odot\lsim 0.8$ (spectral types between M0.5 to M5.5) from 
\citet{downes2014a}. These 15 BDs were spectroscopically confirmed as members of the 
25 Orionis group and its surroundings in Orion OB1a, from an initial sample of 21 BD
candidates. These regions have an age of $\sim7~\rmn{Myr}$, estimated from LMS and BD 
samples \citep{briceno2005,briceno2007a,downes2014a}, all of which are consistent. We 
obtain the disc number fractions of BD and LMS from two different indicators: IR-excesses 
and spectroscopic signatures of ongoing accretion. We compute the fractions on both sides 
of the sub-stellar mass limit following the same procedure, and provide a statistically 
robust treatment that allows us to compute the probability that the disc fractions of 
LMS and BD are different.
%
%
\par
The paper is organized as follows: In Section \ref{observations} we define the sample and describe the 
photometric database, the spectroscopic observations and data reduction. The membership diagnoses are
discussed in Section \ref{memberships}. In Section \ref{discs} we classify the new members as 
class II, evolved discs and class III according to the IR photometric signatures. 
In Section \ref{accretion} we analyse the spectroscopic signatures of accretion and classify 
the new BDs as Classic T Tauri star (CTTS) or Weak T Tauri star (WTTS) sub-stellar analogous. In Section 
\ref{objects} we comment on particular objects and the discussion and conclusions are summarized
in Section \ref{summary}.
%
%
\section{The sample, observations and data reduction}\label{observations}
%
%
For the present work we studied 21 targets from the sample of photometric
BD candidates with expected spectral types between M6 and L1, obtained
during the survey of the 25 Orionis group and its surroundings carried out
by \citet{downes2014a}. The candidate selection was performed based on 
their position in colour-magnitude diagrams that combine I-band optical 
photometry from the CIDA Deep Survey of Orion \citep[CDSO,][]{downes2014a} and 
near IR photometry in the J, H and Ks bands from the Visible and Infrared 
Survey Telescope for Astronomy \citep[VISTA,][]{emerson2004,emerson2010,petrgotzens2011}.
%
%
\par
\begin{figure*}
\includegraphics[width=120mm]{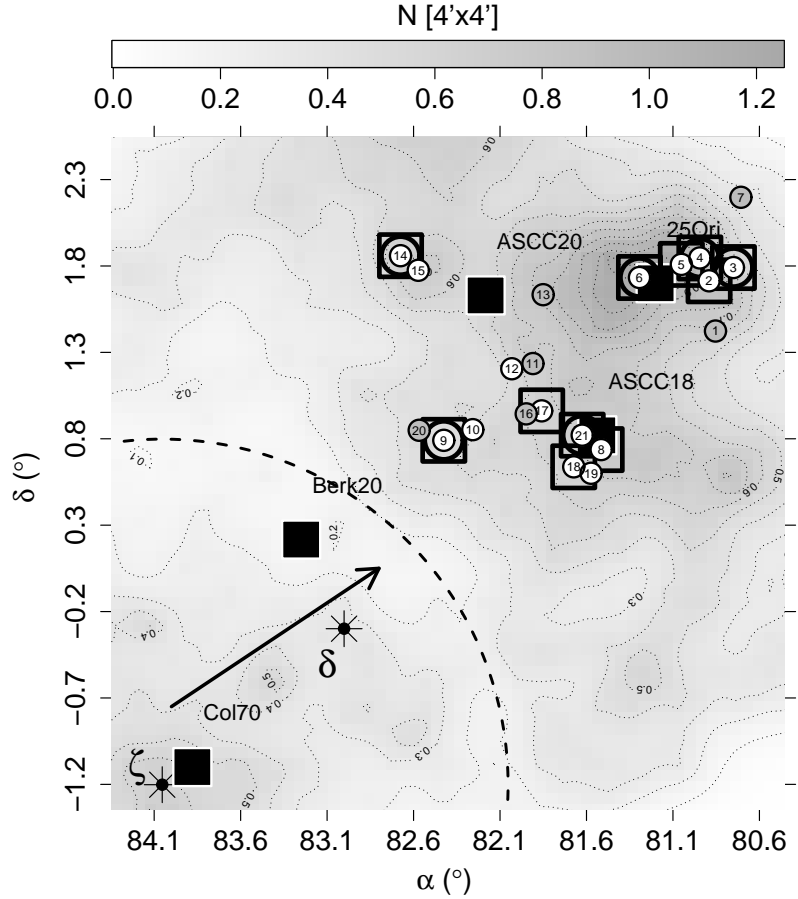}
\caption{Spatial distribution of the sample of the new 15 BDs confirmed in this work 
(labelled white circles) and those confirmed as field dwarfs (labelled gray circles). 
The labelled big asterisks indicate stars of the Orion belt and the dashed circle 
roughly encloses the $\sim5~\rmn{Myr}$ old Orion OB1b sub-association located at 
$\sim440~\rmn{pc}$ from the Sun \citep{briceno2005}. The background gray scale 
indicates the number of sources per $4^\prime\times4^\prime$ bin, of LMS and BD candidates 
from Downes et al. (in preparation) in Orion OB1b, and from \citet{downes2014a} in 
the 25 Orionis group (over-density in the upper right corner) and its surroundings. 
The labelled black squares indicate the known stellar groups in the area from 
\citet{kharchenko2005} together with the 25 Orionis group from \citet{briceno2005}.
The arrow indicates the displacement that a hypothetical star would have if moving from 
Orion OB1b toward the 25 Orionis group during $5~\rmn{Myr}$ with a 
velocity of $2~\rmn{km~s}^{-1}$ (see Section \ref{memberships} for discussion). 
Empty squares indicate the BDs classified as class II or evolved discs and the empty 
circles the CTTS as discussed in Sections \ref{discs} and \ref{accretion}.}
\label{ardec}
\end{figure*}

We have carried out the spectroscopic observations of the 21 candidates with
the OSIRIS instrument \citep{cepa2000} at the Gran Telescopio de Canarias (GTC). 
The 21 candidates analysed in this work were selected as follows: First we
chose 16 candidates with photometric colours consistent with members of
Orion with spectral types M7 to L1. These were randomly selected from the full
catalog of BD candidates, i.e. without imposing any further cuts that might bias
the sample towards objects harbouring discs.
Out of these 16 candidates, 5 objects had another M6 to M7 candidate at a distance
smaller than 7.4$^\prime$, which is close enough to be observed simultaneously with 
the single long-slit configuration of OSIRIS. In this way, we obtained the spectra 
for 21 candidates 
with estimated spectral types between M6 and L1, in only 16 observations.
The spatial distribution of the sample with respect to the Orion OB1 
association is shown in Figure~\ref{ardec}.
%
%
\par
The observations were performed in service mode during March, October and
December 2012 and October and November 2013 as part of the guaranteed Mexican
time with GTC. We used the OSIRIS spectrograph in the single long-slit
configuration with a 1$^{\prime\prime}$ wide slit and the R500R grism,
which results in a $4800\lsim\lambda/${\AA}$\lsim10000$ wavelength coverage with 
a dispersion of 4.88 {\AA}$/pixel$ and a nominal resolution of 587 at 7319 {\AA}. Sky 
flats, dome flats, bias frames and several spectra of comparison lamps for wavelength 
calibration were obtained each night. The observing log is shown in Table \ref{osirislog}.
\begin{table*}
\begin{minipage}{1200mm}
\caption{Observing log for the observations with OSIRIS at GTC}
\label{osirislog}
\begin{tabular}{@{}lccccccc}
\hline
Target\footnote{These IDs allow the identification of the targets in the figures and tables of the article.\\Additional designations from literature are included in the electronic version of Table \ref{photcand}.}  &Beginning of the observation   &$t_{int}$      &Air mass       &Seeing                 &Observing conditions   \\
ID      &[UT]                           &[s]            &               &[$^{\prime\prime}$]    &                       \\
\hline
1       & 2012-03-12T21:02:06.050       & 2400          & 1.26          & 1.5                   & clear/dark \\
2       & 2012-10-08T03:28:22.655       & 2600          & 1.28          & 1.1                   & clear/gray \\
3       & 2012-10-08T04:32:45.348       & 2600          & 1.15          & 1.1                   & clear/gray \\
4       & 2012-12-09T03:15:59.351       & 2600          & 1.26          & 1.0                   & spectroscopic/dark \\
5       & 2012-12-17T01:06:16.672       & 2500          & 1.12          & 1.0                   & photometric/dark \\
6       & 2012-12-17T03:24:48.060       & 2500          & 1.41          & 1.0                   & photometric/dark \\
7       & 2012-12-21T21:25:13.901       & 2600          & 1.61          & 0.8                   & spectroscopic/gray \\
8       & 2013-10-30T06:02:09.175       & 2100          & 1.29          & 0.8                   & clear/dark \\
9       & 2013-10-13T05:02:08.778       & 2500          & 1.13          & 0.7                   & clear/dark \\
10      & 2013-10-13T05:02:08.778       & 2500          & 1.13          & 0.7                   & clear/dark \\
11      & 2013-10-14T05:01:06.925       & 2300          & 1.12          & 0.9                   & photometric/bright \\
12      & 2013-10-14T05:01:06.925       & 2300          & 1.12          & 0.9                   & photometric/bright \\
13      & 2013-11-06T05:05:57.362       & 2300          & 1.20          & 0.7                   & spectroscopic/dark \\
14      & 2013-11-05T04:39:54.909       & 2400          & 1.15          & 0.7                   & spectroscopic/dark \\
15      & 2013-11-05T04:39:54.909       & 2400          & 1.15          & 0.7                   & spectroscopic/dark \\
16      & 2013-11-06T01:50:38.996       & 2300          & 1.26          & 0.8                   & spectroscopic/dark \\
17      & 2013-11-06T01:50:38.996       & 2300          & 1.26          & 0.8                   & spectroscopic/dark \\
18      & 2013-11-06T04:15:15.002       & 2300          & 1.14          & 0.9                   & spectroscopic/dark \\
19      & 2013-11-06T04:15:15.002       & 2300          & 1.14          & 0.9                   & spectroscopic/dark \\
20      & 2013-11-07T02:23:33.637       & 2200          & 1.19          & 0.6                   & spectroscopic/dark \\
21      & 2013-10-30T06:02:09.175       & 2100          & 1.29          & 0.8                   & clear/dark \\
\hline
\end{tabular}
\end{minipage}
\end{table*}
%
%
\par
The spectra were reduced using standard IRAF routines consisting of
bias subtraction, flat-fielding, instrumental response correction, spectrum
extraction, removal of atmospheric spectral features and wavelength calibration.
The wavelength calibration was performed with a mean accuracy of $\sim0.3$ {\AA} 
and cosmic rays were successfully removed during the extraction of the spectra.
%
%
\par
\begin{figure*}
\includegraphics[width=170mm]{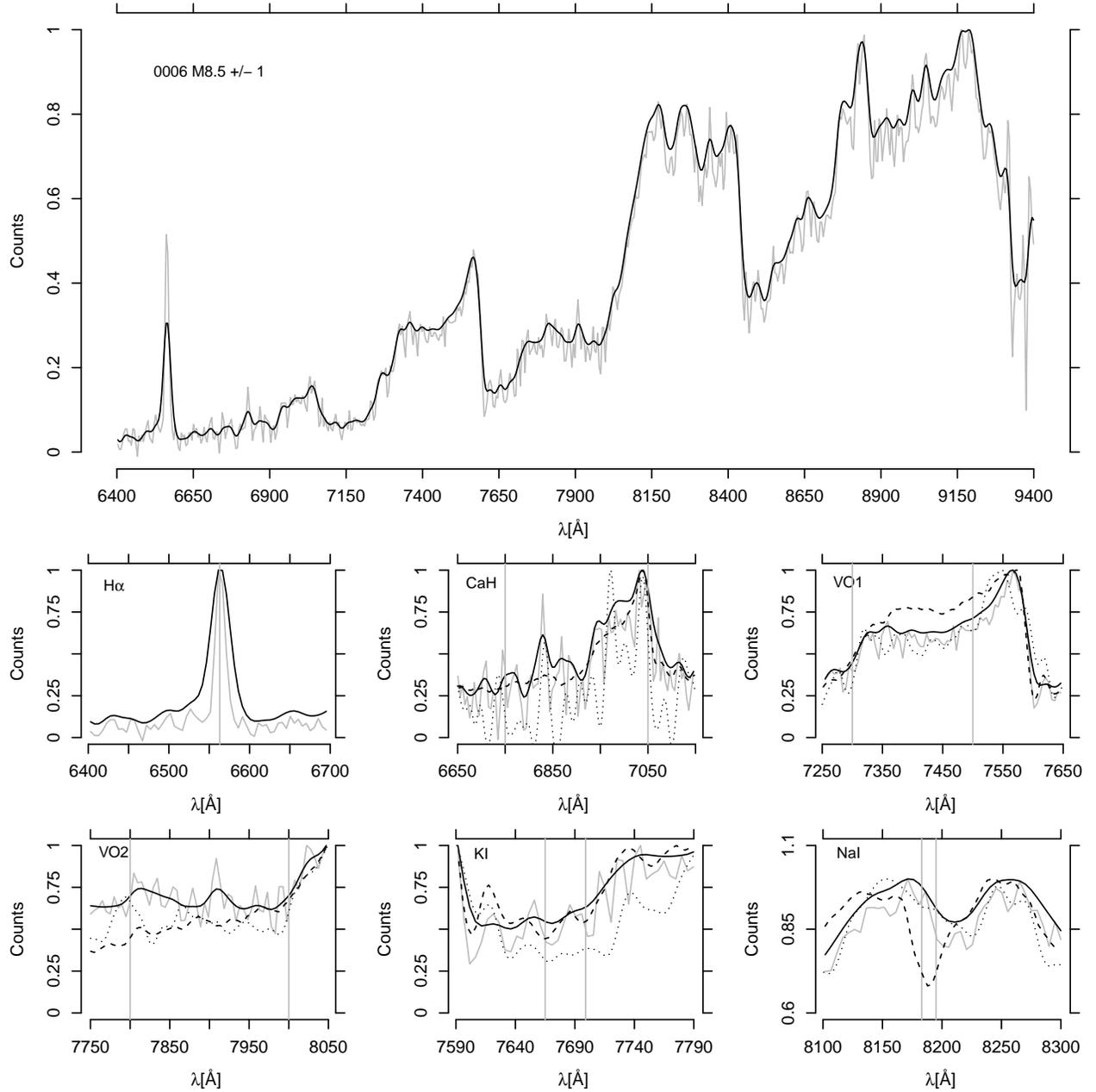}
\caption{Spectra and spectral features for the sample of the young BDs analysed in this work.
Lower pannels show the indicators of youth (VO, NaI, KI and CaI) and ongoing magnetospheric 
accretion (H$\alpha$). Solid lines indicate the spectra for the original resolution 
of $\sim9$ {\AA} (solid gray line) and for $\sim16$ {\AA} resolution (solid black line). 
Dashed and dotted lines indicate respectively a field dwarf of the same spectral 
type from \citet{kirkpatrick1999} and a young dwarf of the same spectral type from 
\citet{luhman2000b}, \citet{briceno2002}, \citet{luhman2003b} and \citet{luhman2004b} 
both at a resolution of $\sim16$ {\AA}. The vertical solid gray lines indicate the 
wavelength range of the corresponding feature. This plot corresponds to the BD 
number 6. The plots for the remaining BDs confirmed as members are available 
in the electronic version of the article.}
\label{spectrum}
\end{figure*}

We computed the spectral types following the semi-automated scheme of
\citet{hernandez2004}\footnote{This procedure was performed using the code SPTCLASS,
available at http://www.astro.lsa.umich.edu/$\sim$hernandj/SPTclass/sptclass.html}
and the equivalent width of the H$\alpha$ line was measured by a linear
fit of the continuum performed with the splot task from IRAF. Figure~\ref{spectrum}
shows the spectra of the new members compared with standards from
\citet{kirkpatrick1999}, \citet{luhman2000b}, \citet{briceno2002}, \citet{luhman2004b}
and \citet{luhman2005} and Table \ref{speccand} shows the derived spectral types,
visual extinctions, equivalent widths of the H$\alpha$ emission line, the 
surface gravity indicators and the final membership diagnosis we will 
explain in Section \ref{memberships}.
\begin{table*}
\label{speccand}
\begin{minipage}{800mm}
\caption{Spectroscopic catalog for the candidates observed with Osiris.}
\begin{tabular}{lcrrrrrrrl}
\hline
Target    &ST             &CaH\footnote{The flags indicate if the spectral feature is consistent with a young BD (1), with a field\\ dwarf (-1) or not conclusive (0). See Section \ref{memberships} for details.}    &VO$^1$ &VO$^2$ &KI     &NaI   &WH$\alpha$     &A$_V$  &Membership     \\
ID        &               &     &       &       &       &       &[{\AA}]        &[mag]  &     \\
\hline
1       & M7.5$\pm$1.0  & -1    & -1    & -1    & -1    & -1    & 0             & 0.37  & field         \\
2       & M9.0$\pm$1.0  & 1     & 1     & 1     & 0     & 1     & -21           & 0.02  & member        \\
3       & M8.0$\pm$1.5  & 0     & 1     & 1     & -1    & 1     & -381          & 0     & member        \\
4       & M6.0$\pm$0.5  & -1    & 1     & -1    & 1     & 1     & -27           & 6.29  & member        \\
5       & M8.0$\pm$1.0  & 1     & 1     & 1     & 0     & 1     & -27           & 0     & member        \\
6       & M8.5$\pm$1.0  & 0     & 1     & 1     & 0     & 1     & -226          & 0     & member        \\
7       & M9.0$\pm$1.5  & -1    & 0     & 0     & 0     & 1     & 0             & 0     & field         \\
8       & M7.5$\pm$2.0  & 0     & 1     & 1     & 0     & 1     & -18           & 0     & member        \\
9       & M8.0$\pm$0.5  & 1     & 1     & 1     & 0     & 1     & -81           & 1.36  & member        \\
10      & M7.0$\pm$0.5  & 1     & 1     & 1     & 1     & 1     & -10           & 0     & member        \\
11      & M7.0$\pm$0.5  & 0     & -1    & 0     & 1     & 0     & -17           & 1.61  & field         \\
12      & M7.5$\pm$0.5  & 1     & 1     & 1     & 0     & 1     & -19           & 0     & member        \\
13      & M9.0$\pm$0.5  & 0     & 0     & 1     & 0     & 1     & 0             & 0     & field         \\
14      & M8.0$\pm$1.5  & -1    & 1     & 1     & 0     & 1     & -50           & 0     & member        \\
15      & M6.0$\pm$0.5  & 1     & 0     & 1     & 1     & 1     & -14           & 0.61  & member        \\
16      & M7.0$\pm$0.5  & -1    & -1    & -1    & 0     & -1    & 0             & 0     & field         \\
17      & M7.0$\pm$0.5  & 1     & 1     & 1     & 1     & 1     & -27           & 0     & member        \\
18      & M8.0$\pm$0.5  & 1     & 1     & 1     & 1     & 1     & -44           & 0     & member        \\
19      & M7.0$\pm$0.5  & 1     & 1     & 1     & 1     & 1     & -18           & 0.21  & member        \\
20      & L1.0$\pm$1.0  & 0     & -1    & 0     & -1    & -1    & 0             &       & field         \\
21      & M7.5$\pm$0.5  & 1     & 1     & 1     & 1     & 1     & -213          & 0.12  & member        \\
\hline
\end{tabular}
\end{minipage}
\medskip
\end{table*}
%
%
\par
In order to detect possible IR-excesses from the spectral energy distribution (SED),
we complemented the I-band photometry from CDSO and the J, Z, Y, H and Ks-band
photometry from VISTA with additional photometry at $3.6$, $4.5$, $5.8$ and 
$8.0~\mu\rmn{m}$ from IRAC-Spitzer observations from \citet{hernandez2007} and
Brice\~no et al. (in preparation), and at $3.4$, $4.6$, $12$ and $22~\mu\rmn{m}$ 
from the WISE All-Sky Source Catalog \citep{wright2010}. Additional photometric data in 
the g, r, i and z-bands from the Sloan Digital Sky Survey Catalog Data Release 8 
\citep{adelmanMcCarthy2011} were obtained using the VizieR virtual observatory 
system \citep{ochsenbein2000}. Because of the different sensitivity limits and 
spatial coverage of each survey, not all the photometric bands are available for all 
the candidates. The available photometric information for the new BDs is summarized 
in Table \ref{photcand}.
\begin{center}
\begin{table*}
\begin{minipage}{800mm}
\caption{Photometric catalog of the new BDs confirmed as members of 25 Ori or 
Orion OB1a}
\label{photcand}
\begin{tabular}{lcccccccc}
\hline
ID \footnote{The complete version of the table is available in the electronic version of the article.} &     RA&              DEC&            I&      J&      Z&      K&      $3.6~\mu\rmn{m}$&      $4.5~\mu\rmn{m}$      \\
\hline
2&      80.791100&       1.714128&       20.79&  17.49&  19.77&  16.42&  16.27&          16.72 \\
3&      80.850308&       1.790936&       19.93&  17.26&  19.41&  16.21&  15.68&          15.00 \\
4&      80.894376&       1.846516&       19.79&  16.52&  18.94&  15.12&  14.80&          14.31 \\
5&      80.951425&       1.809198&       20.20&  17.30&  19.49&  16.23&  15.98&          15.62 \\
6&      81.243647&       1.733360&       20.32&  17.35&  19.49&  16.24&  16.07&          15.65 \\
8&      81.514956&       0.737391&       19.25&  16.67&  18.39&  15.84&  15.43&          15.20 \\
9&      82.375058&       0.790957&       20.76&  17.42&  19.58&  16.32&  16.02&          15.71 \\
10&     82.307508&       0.852826&       18.45&  16.13&  17.83&  15.26&  15.09&          14.88 \\
12&     81.956878&       1.205257&       18.35&  16.21&  17.70&  15.38&  15.14&          15.04 \\
14&     82.675353&       1.859462&       20.21&  17.41&  19.57&  16.33&  16.07&          15.71 \\
15&     82.647725&       1.775001&       18.15&  16.02&  17.41&  15.22&  14.98&          14.77 \\
17&     81.909127&       0.961591&       18.26&  15.99&  17.77&  15.09&  14.72&          14.02 \\
18&     81.673711&       0.637468&       19.75&  17.04&  19.18&  16.03&  15.51&          15.02 \\
19&     81.648441&       0.600469&       18.70&  16.22&  17.96&  15.38&  15.10&          14.83 \\
21&     81.549774&       0.820961&       19.46&  16.67&  18.52&  15.73&  15.51&          14.89 \\
\hline
\end{tabular}
\end{minipage}
\medskip
\end{table*}
\end{center}
%
%
\par
As discussed by \citet{jarrett2011} there is an offset between WISE $3.6~\mu\rmn{m}$ and 
IRAC $3.4~\mu\rmn{m}$ magnitudes, as well as between WISE $4.5~\mu\rmn{m}$ and IRAC $4.6~\mu\rmn{m}$ 
magnitudes. Both offsets occur approximately for $m_{3.6~\mu\rmn{m}}>14$ magnitudes and
$m_{4.5~\mu\rmn{m}}>13$ magnitudes in such a way that the magnitudes from WISE are 
increasingly fainter than the corresponding from Spitzer. According to 
\citet{jarrett2011} these biases occur as a consequence of an overestimation in the 
background levels of the WISE images. We found exactly the same behaviour affecting 
all of our faint objects with available photometry from WISE and Spitzer. In order 
to correct the photometry from WISE, we selected $\sim10000$ objects in the Spitzer 
fields with available photometry from WISE and fit the residuals of the corresponding 
magnitudes as a function of one of them. After that, we applied the corresponding 
offset to the WISE photometry as a correction of the reported magnitudes. Because 
IRAC-Spitzer has no measurements around $12~\mu\rmn{m}$ it was not possible 
to establish if the background in the WISE $12~\mu\rmn{m}$ pass-band is also underestimated. 
Only two BDs (numbers 4 and 18) have measurements in the WISE $12~\mu\rmn{m}$ pass-band
with reasonable SNR. We disregard all the measurements in the WISE $22~\mu\rmn{m}$ pass-bands 
because of the very low SNR.
%
%
\section{Membership diagnoses}\label{memberships}
Young BDs are still contracting and their surface gravity is lower than for an old field
dwarf of the same effective temperature. Since the contamination of the candidate sample
comes from old field dwarfs \citep{downes2014a}, we evaluate the membership of the
candidates to the 25 Orionis group or Orion OB1a, according to several spectral features
sensitive to surface gravity following \citet{mcgovern2004}.
We considered the following spectral features: $(i)$ The CaH ($\lambda 6750$ to $\lambda 7050$) molecular band, which is stronger in field dwarfs; $(ii)$ the VO$^1$ and VO$^2$ molecular bands ($\lambda 7300$ to $\lambda 7500$ and $\lambda 7800$ to $\lambda 8000$ respectively), which are weaker in field dwarfs; $(iii)$ the atomic lines KI $\lambda 7665,7699$ and NaI $\lambda 8183,8195$, which are expected to be stronger in field dwarfs. Other spectral features sensitive to surface gravity, such as the absorption lines RbI $\lambda 7800,7948$ and CsI $\lambda 8521$, were marginally detected only in some of the spectra because of the low SNR and wavelength resolution and did not allow for a reliable estimation of the surface gravity, hence they were not considered further.

These 5 spectral features were compared in the spectra of our candidates to
those of young dwarfs from \citet{luhman2000b},
\citet{briceno2002}, \citet{luhman2003b} and \citet{luhman2004b}, and field dwarfs from
\citet{kirkpatrick1999} of the same spectral types. 
We found a very good general agreement between the different surface gravity indicators as we show in Table \ref{speccand} and Figure~\ref{spectrum}. 
Table \ref{speccand} summarizes, for each of the 21 candidates, whether each of the 5 spectral features is consistent with a young or field BD of the corresponding spectral type.
We classified as members those candidates showing at least 3 spectral features consistent
with low surface gravity.
Using these criteria we have confirmed 15 members from the sample of 21 photometric
candidates.

\par 
We stress that such a selection is also supported by: $(i)$ all the objects classified as
young BDs show H$\alpha$ line in emission and $(ii)$ their visual extinction is consistent 
with the mean value for known members in 25 Ori and Orion OB1a 
\citep[$A_V\sim0.4\pm0.3$][]{downes2014a,briceno2007a,briceno2005}. The visual extinction 
for each candidate was obtained from the observed I-J colour and the intrinsic I-J colour that 
corresponds to the spectral type. We used the intrinsic colours to spectral type relationships 
used by \citet{luhman1999}, \citet{briceno2002} and \citet{luhman2003b}, designed to match the 
\citet{baraffe1998} tracks such that the components of GG Tau appear coeval on the H-R diagram, 
as explained by \citet{luhman2003b}. We used the extinction law from \citet{cardelli1989} assuming 
R$_V=$ 3.09. The resulting extinctions are shown in Table \ref{speccand} and are consistent,
within the uncertainties, with previous determinations. The only 
exceptions are for the BDs 4 and 9 and are discussed in Section \ref{objects}.
%
%
\par
The selected young BDs could be members of the 25 Orionis group or other regions of Orion OB1a which essentially 
have the same age \citep[$\sim7~\rmn{Myr}$,][]{briceno2005,downes2014a}. Nevertheless, the BDs could 
also be members of a different sub-region within the Orion OB1 association. Particularly we are 
interested in determining if some of them could belong to the younger and relatively close 
($\sim3^\circ$, Figure \ref{ardec}) Orion OB1b \citep[$\sim5~\rmn{Myr}$,][]{briceno2005}. However, we conclude 
the contamination due to members of the nearby younger Orion OB1b must be very low for the following reasons: 

$(i)$ The spatial distribution of the new BDs is consistent with those from previously known members 
of the 25 Orionis group and Orion OB1a of earlier spectral types \citep{briceno2005,downes2014a,kharchenko2005}. 
As shown in the spatial overdensities of Figure~\ref{ardec} the candidate LMS and BDs of Orion OB1a and 
the 25 Orionis group are spatially distinct from the region populated by Orion OB1b sources.

$(ii)$ Considering a velocity dispersion of $\sim2~\rmn{km~s}^{-1}$, the angular separation between the sub-association 
Orion OB1a and OB1b and their distances to the Sun ($\sim360~\rmn{pc}$ for OB1a \citep{briceno2007a} and $\sim440~\rmn{pc}$ 
for OB1b \citep{briceno2005}), a member of the younger OB1b \citep[$\sim5~\rmn{Myr}$,][]{briceno2005} would need more 
than $5~\rmn{Myr}$ to escape and end up in the line of sight to the 25 Orionis group. Thus, the possible contaminants could, 
at most, affect the eastern part of the survey. However, if the contamination from OB1b were significant an overdensity 
of class II, evolved discs and/or accretors (see Sections \ref{discs} and \ref{accretion}) would be expected in the 
eastern part of the survey closer to Orion OB1b, which is not observed in Figure~\ref{ardec}. Therefore, we consider 
it much more likely that these young BDs are members of the 25 Ori group or the Orion OB1a sub-association, than 
Orion OB1b contaminants.

$(iii)$ A particular case is the new BDs 8, 18, 19 and 21 which are close to the stellar group ASCC18 
from the catalog of \citet{kharchenko2005} who report it as a group in the Orion OB1 association. On the basis 
of one intermediate mass star, \citet{kharchenko2005} estimated an age of $\sim13~\rmn{Myr}$ for ASCC18 which is 
slightly older than the 25 Orionis group and Orion OB1a, something which does not affect the conclusions presented here.
%
%
\par
\begin{figure*}
\includegraphics[width=120mm]{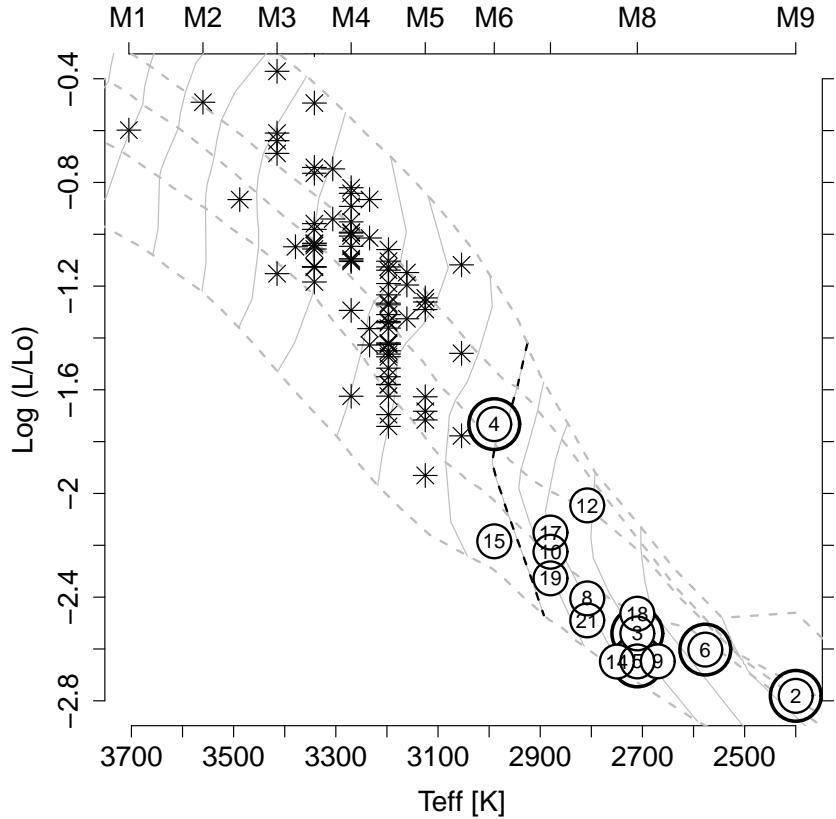}
\caption{H-R diagram of the new young BDs confirmed in this work (labelled circles)
together with LMS from \citet{downes2014a} (asterisks). The big circles indicate the
BDs placed in the 25 Orionis group. The evolutionary tracks for $0.8$, $0.7$, 
$0.6$, $0.5$, $0.4$, $0.3$, $0.2$, $0.1$, $0.072$ (dashed), $0.06$, $0.05$, $0.04$, 
$0.03$, $0.02$ and $0.01~\rmn{M}_\odot$, and isochrones for $1$, $3$, $5$, $10$ and 
$30~\rmn{Myr}$ from the DUSTY models of \citet{chabrier2000} are also indicated. 
We applied a fiducial $40~K$ offset in temperature to the BDs 9 and 14 for clarity.}
\label{hr}
\end{figure*}

The position of the new BDs in the H-R diagram is shown in Figure~\ref{hr} together
with the LMS of the 25 Orionis group from \citep{downes2014a}. We computed the
effective temperatures by interpolation of the spectral types into the \citet{luhman2003b}
relationships for young BDs and we computed the bolometric luminosities from the dereddened I-band
magnitudes, assuming a distance of $360~\rmn{pc}$ \citep{briceno2007a} and the bolometric corrections
from \citet{dahn2002}. The masses were derived by interpolation of the luminosities 
and the effective temperatures into the DUSTY models from \citet{chabrier2000} and are 
shown in Table \ref{physprop}.
%
%
\par
Finally, based on simulations performed with the Besan\c{c}on Galactic model
\citep{robin2003}, we showed in \citet{downes2014a} that in the spectral-type
range considered here, our photometric selection procedure for candidates in
the 25 Orionis group and its surroundings has a mean total
contamination of $\sim25~\rmn{\%}$, composed solely of foreground field dwarfs.
In this work we spectroscopically confirmed that $71.4_{-9.0}^{+8.0}~\rmn{\%}$ of
the candidates are real members and the remaining targets show spectral features 
consistent with the high surface gravity expected from old dwarf stars from the 
field, which perfectly matches the expected contamination fraction. 
%
%
\section{Photometric signatures of discs}\label{discs}
%
%
It is well established that some young BDs show IR excesses indicative of
circumstelar discs \cite[e.g.][]{luhman2012}. In this section we classify
the discs surrounding the new BDs according to their excesses in the
wavelength range $\lambda>2~\mu\rmn{m}$.
%
%
For the classification of the discs we assume the general scheme for their evolution 
around LMS and VLMS \citep[e.g.][]{lada2006} in which they evolve from class 0 to I, 
then to class II, then passing through the evolved disc stage \citep[e.g.][]{hernandez2007}, 
perhaps passing through the pre-transitional to the transitional stages \citep[e.g.][]{espaillat2007}
depending on the mass of the disc, and ending as class III objects. Thus, we follow the same 
classification scheme in order to compare the evolution of discs around objects at 
both sides of the sub-stellar mass limit.
%
%
\par
The evolutionary stage of the discs can be inferred from the slopes of
the spectral energy distributions (SED) at different wavelength ranges
or equivalently from its distribution in selected colour-colour diagrams.
%
%
We performed the classification according to the SEDs of each new member and 
plotted the results in a set of selected colour-colour diagrams in order to 
show the distribution of the BDs and how it supports their classification 
through the SEDs. All the SEDs include measurements at the I, Z, Y, J, H, Ks, 
$3.4~\mu\rmn{m}$ and $4.6~\mu\rmn{m}$ photometric-bands, although the differences in sensitivity 
and spatial coverage of the WISE, IRAC and SDSS surveys give us information only
for a subset of the complete sample in the remaining photometric-bands mentioned
in Section \ref{observations}.
%
%
\par
We consider as IR excesses all the SED points for $\lambda\gsim2~\mu\rmn{m}$ showing
fluxes above the photospheric emissions predicted by the BT-Dusty model
from \citet{allard2012} which was developed for near-IR studies of BDs with 
$T_{eff} > 1700 K$. Our procedure was as follows:
%
%
First, we estimate the photospheric emission for the BDs by fitting their 
extinction corrected SEDs to the BT-Dusty model.
For all the fits we used the I, Z, Y, J and H band-passes which were complemented
with the g, r, i and z photometry when available. In this way, we perform the fits
with a minimum of 5 and a maximum of 10 band-passes in a wavelength range where
the IR excesses are not expected to occur and where the peaks of the photospheric
contribution to the SEDs are expected. Before the fits, the SEDs were corrected
by reddening using the A$_V$ obtained as explained in Section \ref{memberships} and 
the extinction law from \citet{fitzpatrick1999} improved by \citet{indebetouw2005} 
in the infrared ($\lambda>1.2~\mu\rmn{m}$), with the parameter R=3.02.
We make use of the Virtual Observatory SED Analyzer (VOSA) fitting tools
from \citet{bayo2008} in order to apply the reddening corrections and make
the fits to the \citet{allard2012} models. The effective temperatures
obtained from the fits are consistent with those obtained by the interpolation 
of the spectral types in the \citet{luhman2003b} relationships.
%
%
\par
We classify as class II those BDs showing excesses at the K-band or longer
wavelengths, consistent with the median SED for Taurus from \citet{furlan2006}.
We considered as evolved discs those BDs showing excesses for wavelengths
longer than the Ks-band but clearly weaker than the excesses observed for 
class II. Finally we classify as BDs of class III those showing a purely 
photospheric SED consistent with the fits to the \citet{allard2012} models.
%
%
\par
\begin{figure*}
\label{sed}
\includegraphics[width=180mm]{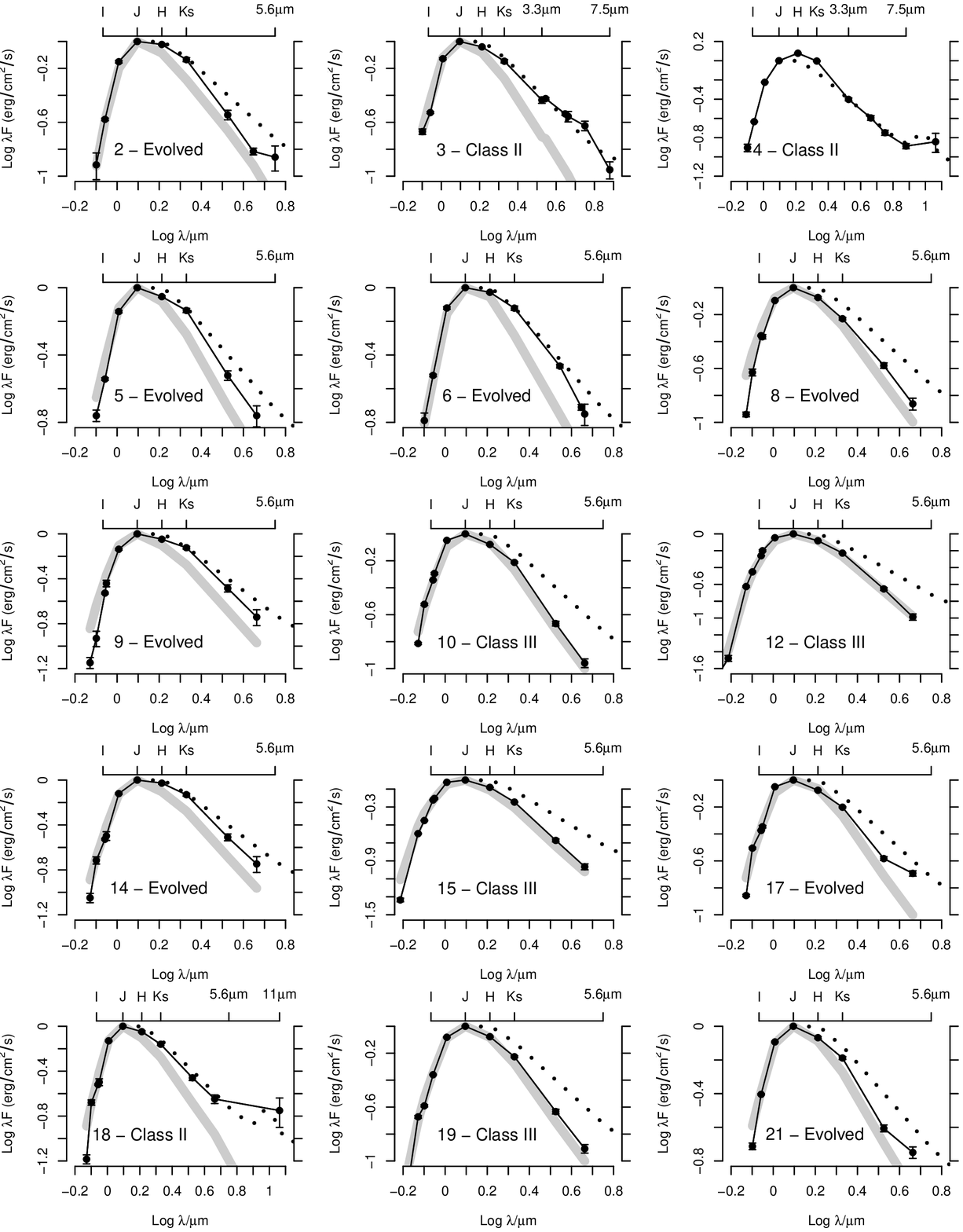}
\caption{The black dots and the black solid lines indicate the observed SEDs of the new BDs. 
The gray lines show the photospheric SED that results from the interpolation of the 
fluxes in the BT-Dusty models from \citet{allard2012} and the dotted lines indicate 
the median SED for Taurus from \citet{furlan2006}. All the SEDs are normalized to 
the flux at the J-band. Each pannel shows the ID of the BD and its classification as 
class II, evolved disc or class III. The BD number 4 shows clear excesses for H-band 
and longer wavelengths and the number of points at shorter wavelengths does not allow 
for a reliable fit into the photospheric models.}
\end{figure*}

With the available photometry we cannot reliably detect BDs surrounded by
transitional or pre-transitional discs \citep{espaillat2007} because the spectral
coverage in the wavelength range $8\lsim\lambda/\mu\rmn{m}\lsim 11$ is not homogeneous 
for most of the sources. Figure~\ref{sed} shows the SEDs for all the new confirmed 
BDs, together with the resulting fit to the photospheric models from \citet{allard2012}
and the median SED for Taurus from \citet{furlan2006}.
%
%
\par
\begin{figure}
\includegraphics[width=80mm]{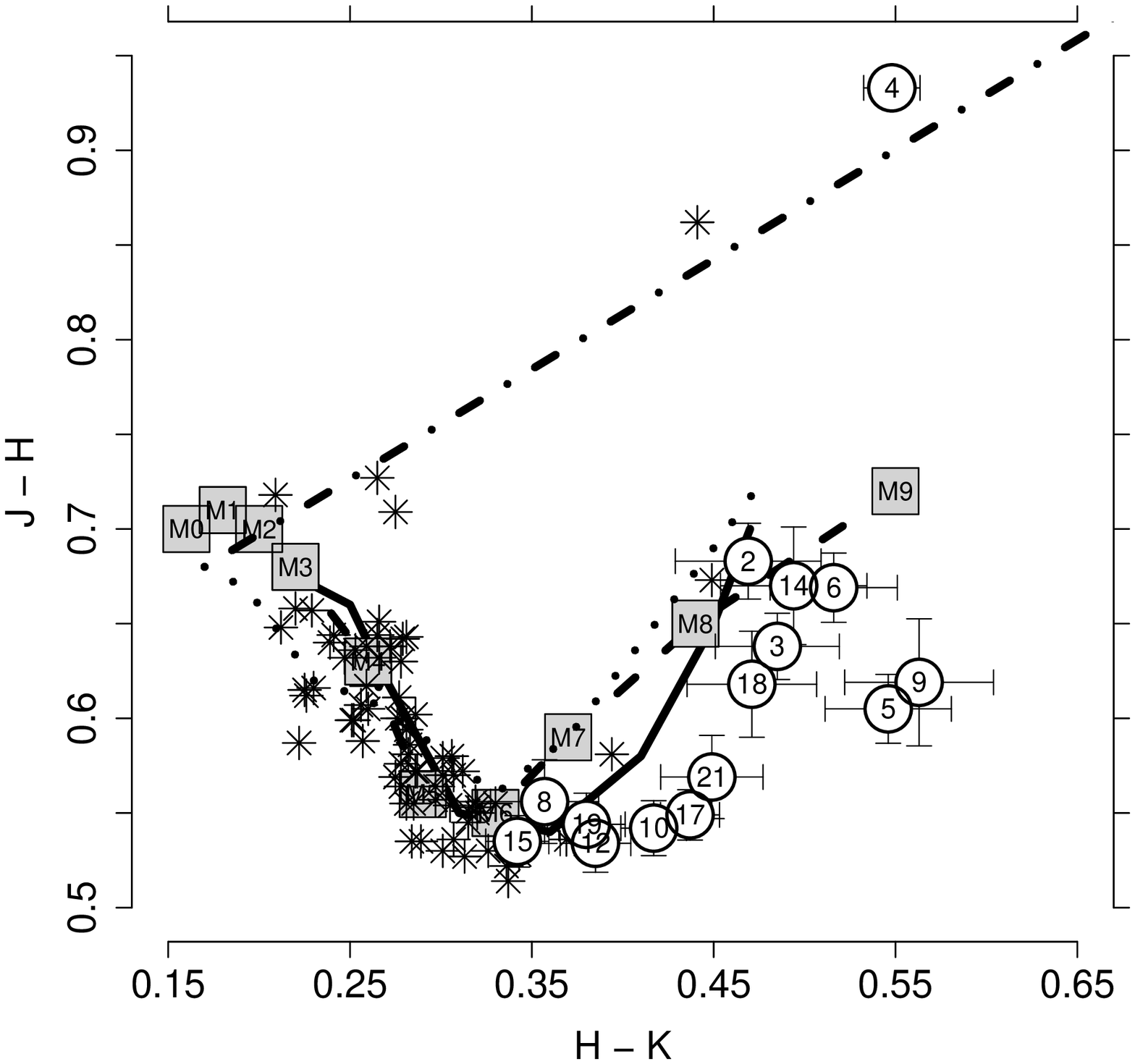}
\includegraphics[width=80mm]{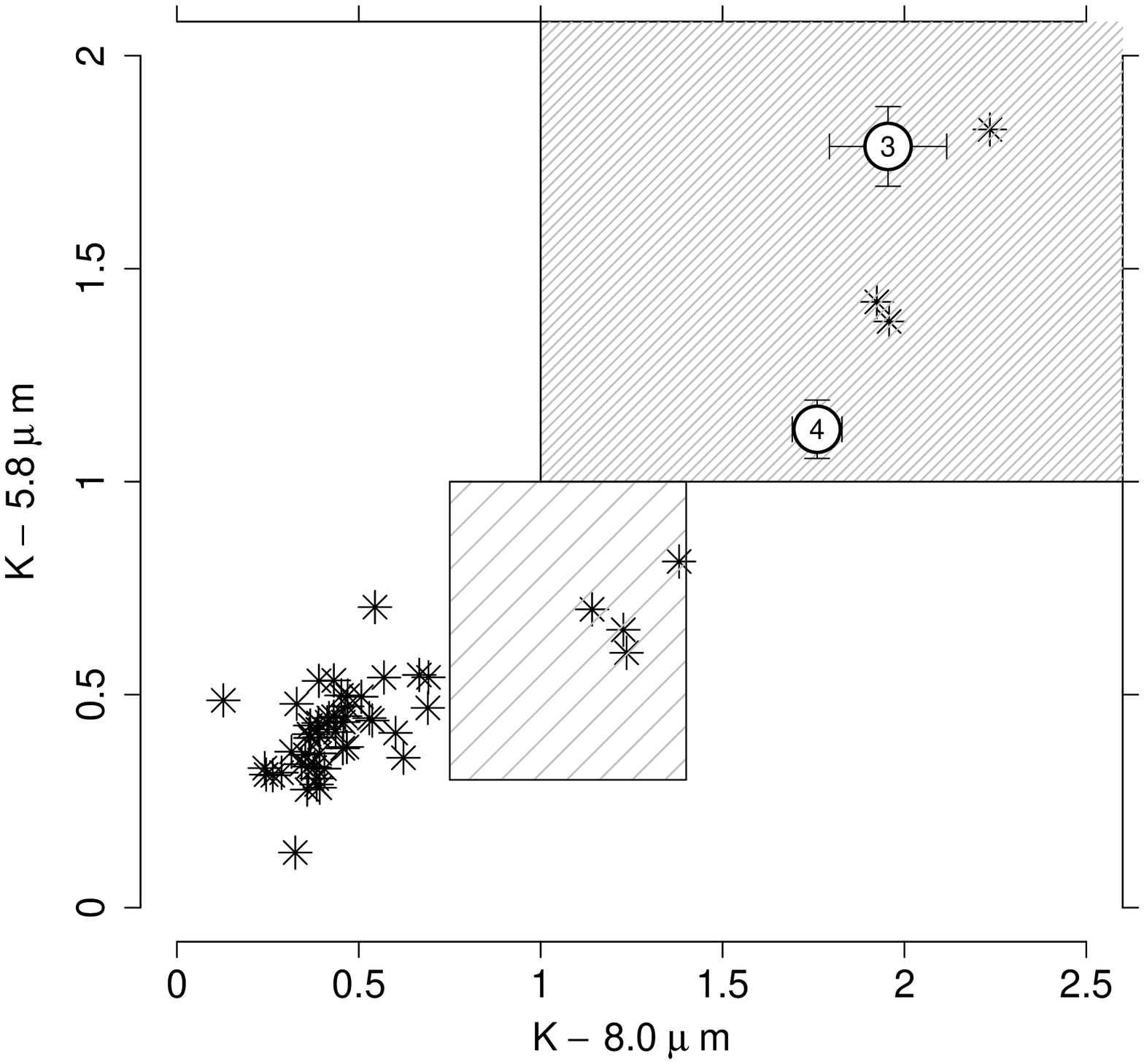}
\caption{Distribution of the new BDs (labelled circles) and LMS from \citet{downes2014a} 
(asterisks) in colour-colour diagrams. The upper panel shows the observed J-H vs. H-K diagram 
including three photospheric locii: from \citet{luhman1999}, \citet{briceno2002} and 
\citet{luhman2003b} (dotted line);  from \citet{luhman2010} (dashed line with labeled squares 
indicating spectral types) and from \citet{pecaut2013} (solid line). The dash-dotted line indicates 
the CTTS locus from \citet{meyer1997} for M0 stars. In the lower panel the shadowed polygon and the 
dashed polygon indicate respectively disc and transitional disc loci computed from the colour intervals 
defined by \citet{luhman2010}. Only BDs 3 and 4 have information in the $8~\mu\rmn{m}$ passband.}
\label{cc1}
\end{figure}
\begin{figure}
\includegraphics[width=80mm]{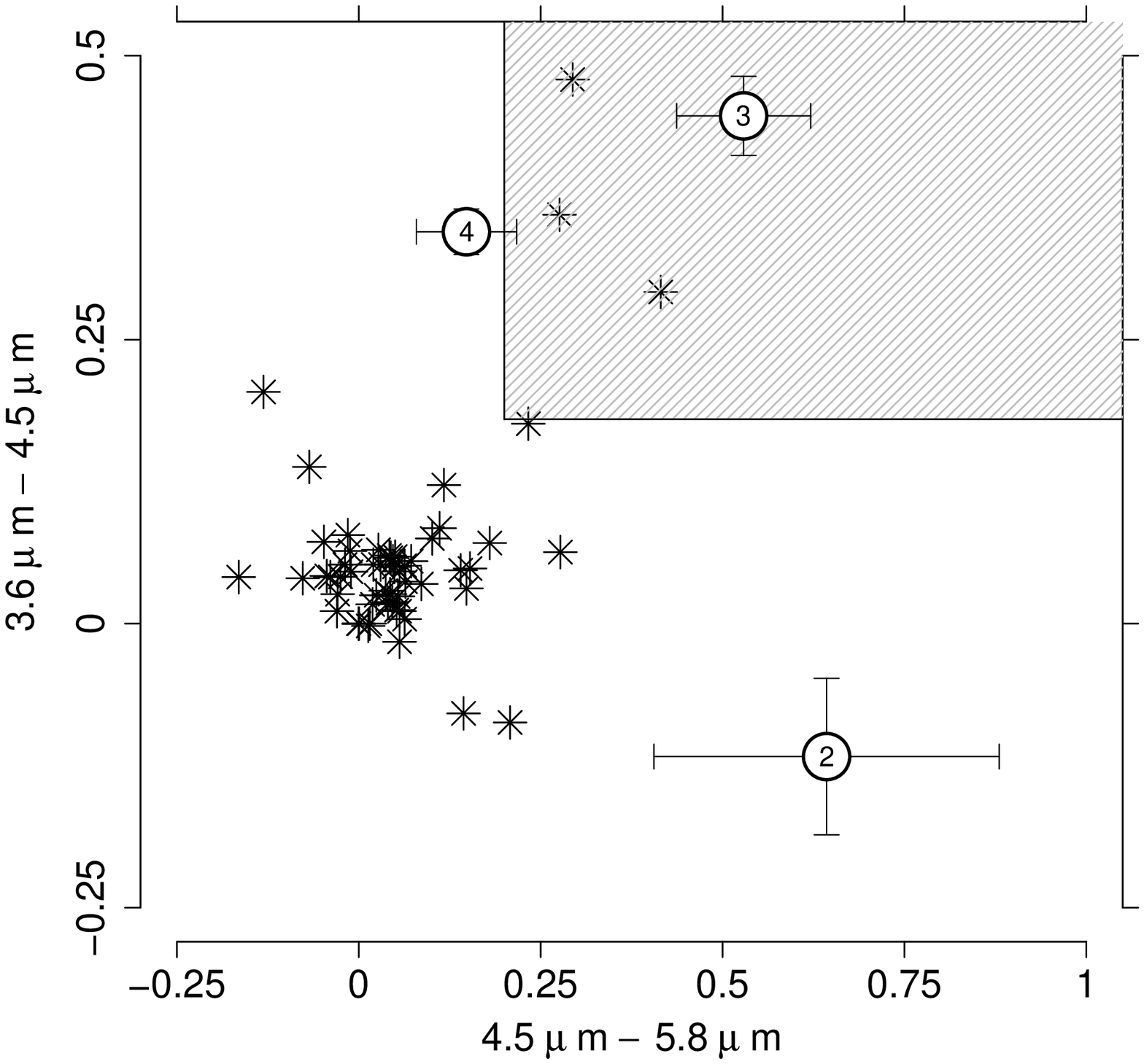}
\includegraphics[width=80mm]{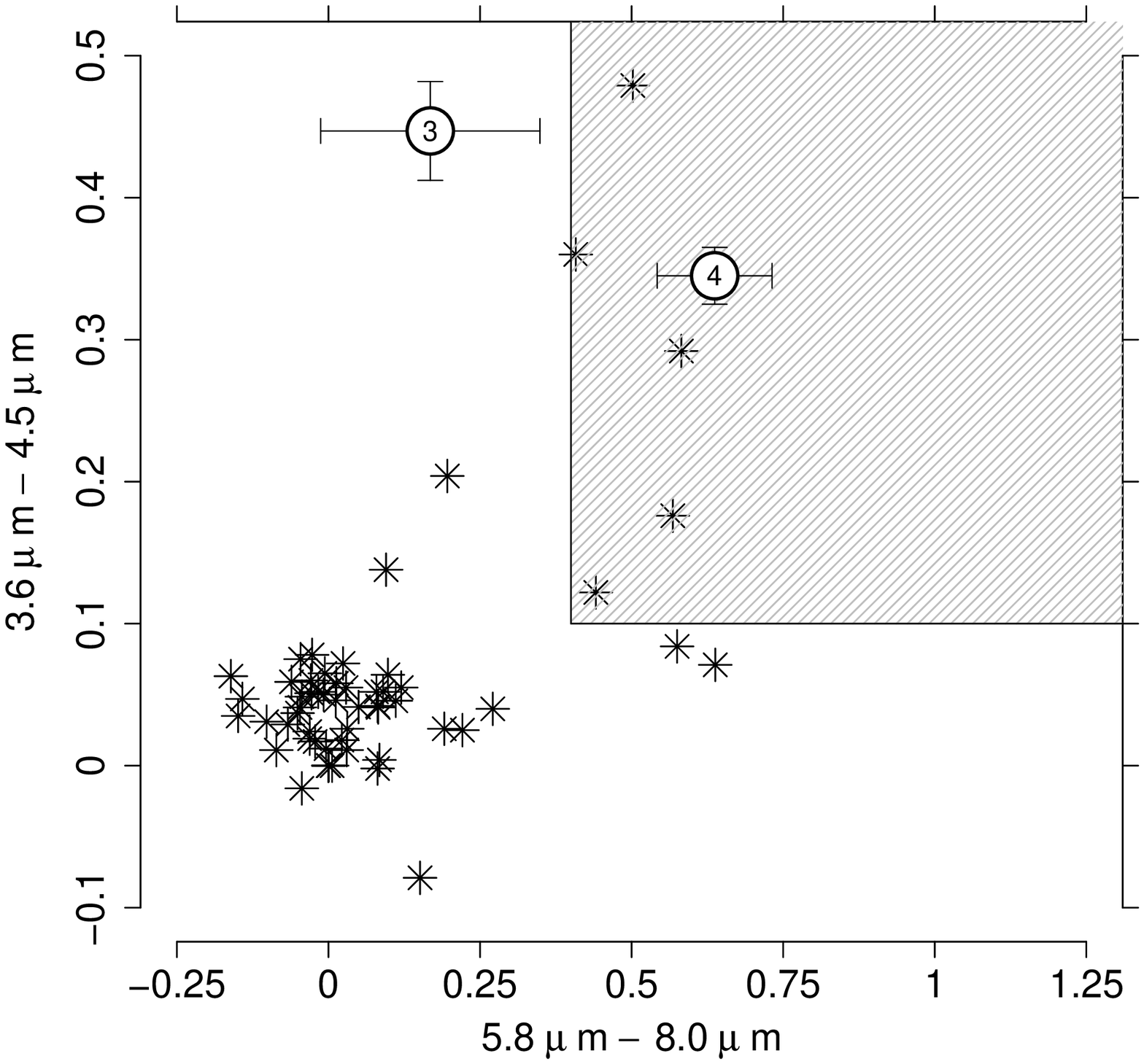}
\caption{Distribution of BDs in colour-colour diagrams.
Symbols are as in Figure~\ref{cc1}. The shadowed areas indicate
the CTTS locus from \citet{hartmann2005a} (upper panel), 
and the excess region defined by \citet{luhman2005} (lower panel).
Only BDs 2, 3 and 4 have information for the $5.8~\mu\rmn{m}$
passband.}
\label{cc2}
\end{figure}

The IR excesses are also detectable in colour-colour diagrams
supporting the results obtained from the SEDs. 
Figures \ref{cc1} and \ref{cc2} show a selection of colour-colour diagrams 
some of those showing the discs locii from \citet{hartmann2005a}, \citet{luhman2005}
and \citet{luhman2010}. 
%
%
\par
Summarizing the results from our classification, out of the 15 new BDs, 
we have found 3 objects showing IR excesses consistent with BDs of class II, 
8 consistent with evolved discs and 4 showing purely photospheric SEDs that 
were classified as BDs of class III.
%
%
Using data from \citet{downes2014a}, we have followed the same procedure to 
classify IR excesses observed in a sample of 77 LMS with spectral types 
between M0.5 and M5.5 and masses $0.1\lsim \rmn{M}/\rmn{M}_\odot\lsim0.8$, 
confirmed as members of the 25 Orionis group. We find 3 of class II, 11 having 
evolved discs and 63 of class III. These results are summarized in Table \ref{nir}. 
%
%
\par
An important issue is that our selection of the photometric candidates was performed inside 
a set of locii in the colour-magnitude diagrams I vs. I-J, I vs. I-H and I vs. I-K that enclose 
all the LMS and BD expected to have purely photospheric emissions as well as those showing IR 
excesses \citep{downes2014a}. Then, our procedure do not bias the selection of candidates towards 
LMS or BDs harbouring discs, even for those candidates that are fainter than the completeness limits 
in such diagrams \citep{downes2014a}. 
\begin{table*}
\label{nir}
\caption{Number fraction of BD and LMS classified as class II, evolved disc, class III, WTTS and CTTS.}
\begin{tabular}{llllrc}
\hline
Classification               & \multicolumn{2}{c}{BD ($\lsim 0.072~\rmn{M}_\odot$)} & \multicolumn{2}{c}{LMS ($0.1\lsim \rmn{M}/\rmn{M}_\odot\lsim0.8$)} & P$_{>0.1}$ \\
                             &[N]      & [\%]                                 & [N] & [\%]                  &             \\
\hline									
Class II                     &3        & $20.0^{+9.9}_{-7.9}$                 &3    & $3.9^{+2.4}_{-1.6}$   & $0.9908$    \\
Evolved                      &8        & $53.3^{+10.9}_{-11.0}$               &11   & $14.3^{+3.8}_{-3.2}$  & $0.9987$    \\
CTTS               	     &5        & $33.3^{+10.8}_{-9.8}$                &3    & $3.9^{+2.4}_{-1.6}$   & $0.9993$    \\
CTTS \& Class II or Evolved  &5        & $33.3^{+10.8}_{-9.8}$                &3    & $3.9^{+2.4}_{-1.6}$   & $0.9993$    \\
Class III                    &4        & $26.7^{+10.4}_{-9.0}$                &63   & $81.8^{+3.8}_{-4.0}$  & $0.9988$    \\
WTTS                         &10       & $66.7^{+9.7}_{-10.9}$                &74   & $96.1^{+1.7}_{-2.3}$  & $0.9809$    \\
\hline                               
\end{tabular}
\medskip
\end{table*}
\section{Computation of disc fractions}\label{fractions}
%
%
In order to compute discs fractions and their uncertainties given the
limited number of objects with discs in our samples, we use Bayesian
statistics.
%
%
This allows us to state the problem in a general way providing us with
a probability function for the disc fractions and, more importantly, with a
quantitative way of estimating the probability that disc fractions from
two independent samples differ.
\par
%
%
We can think of the disc fraction $f_{d}$ as the probability for any
single star to harbour a disc. Given a sample with a disc fraction
$f_{d}$ and a total number of $N_T$ stars, the likelihood
$P(N_{d},N_{T}|f_{d})$ of observing $N_{d}$ stars with discs
is then simply given by the Binomial distribution.
Assuming a uniform prior probability distribution for the disc fraction 
in the range $0\leq f_{d}\leq 1$, from the Bayes's theorem \citep{sivia2006} 
we have the Posterior PDF expressed simply as:
\begin{equation}\label{ec:posterior1d}
P(f_{d}|N_{d},N_{T})=Cf_{d}^{N_{d}}(1-f_{d})^{N_T-N_{d}}
\end{equation}
where $C$ is a normalization constant such that $\int df_{d} P(f_{d}|N_{d},N_{T})=1$.
%
%
\par
We can now generalize this to the case where we have two independent samples,
LMS and BDs, and express the full posterior probability of $f_{d}^{V}$
\emph{and} $f_{d}^{B}$ as the product of the two independent probabilities:
\begin{equation}\label{ec:posterior_vl_bd}
P(f_{d}^{V},f_{d}^{B}|N_{d}^{V},N_{T}^{V},N_{d}^{B},N_{T}^{B})= P(f_{d}^{V}|N_{d}^{V},N_{T}^{V})P(f_{d}^{B}|N_{d}^{B},N_{T}^{B})
\end{equation}
where $N_{d}^{V}$ and $N_{d}^{B}$ are the observed numbers of LMS
and BDs harbouring discs, in samples with a total number of $N_{T}^{V}$
and $N_{T}^{B}$ of LMS and BDs respectively. Each of the terms in the right
hand side in this equation are thus given by Eq. \ref{ec:posterior1d}.
\par
\begin{figure}
\includegraphics[width=90mm]{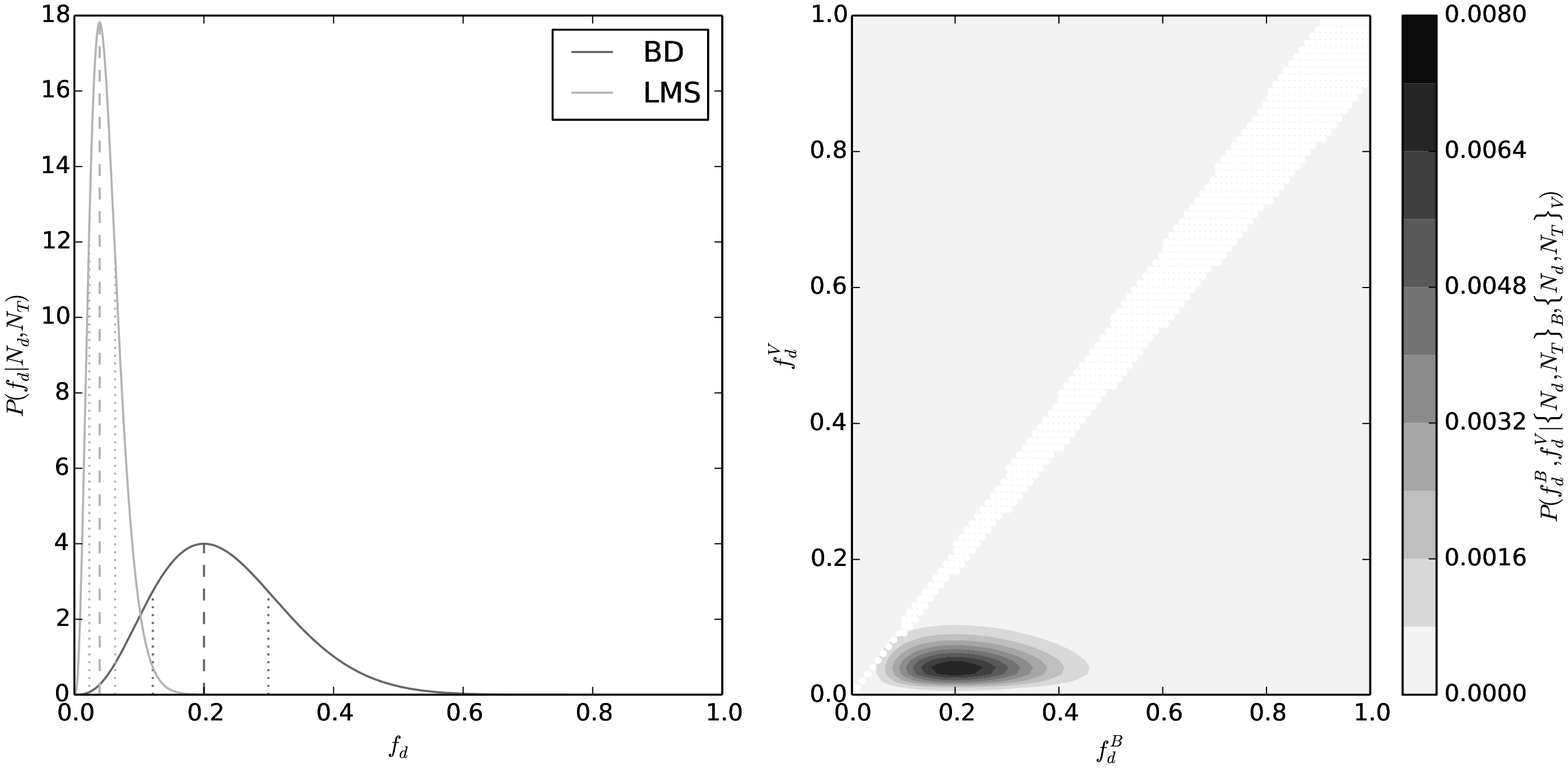}
\includegraphics[width=90mm]{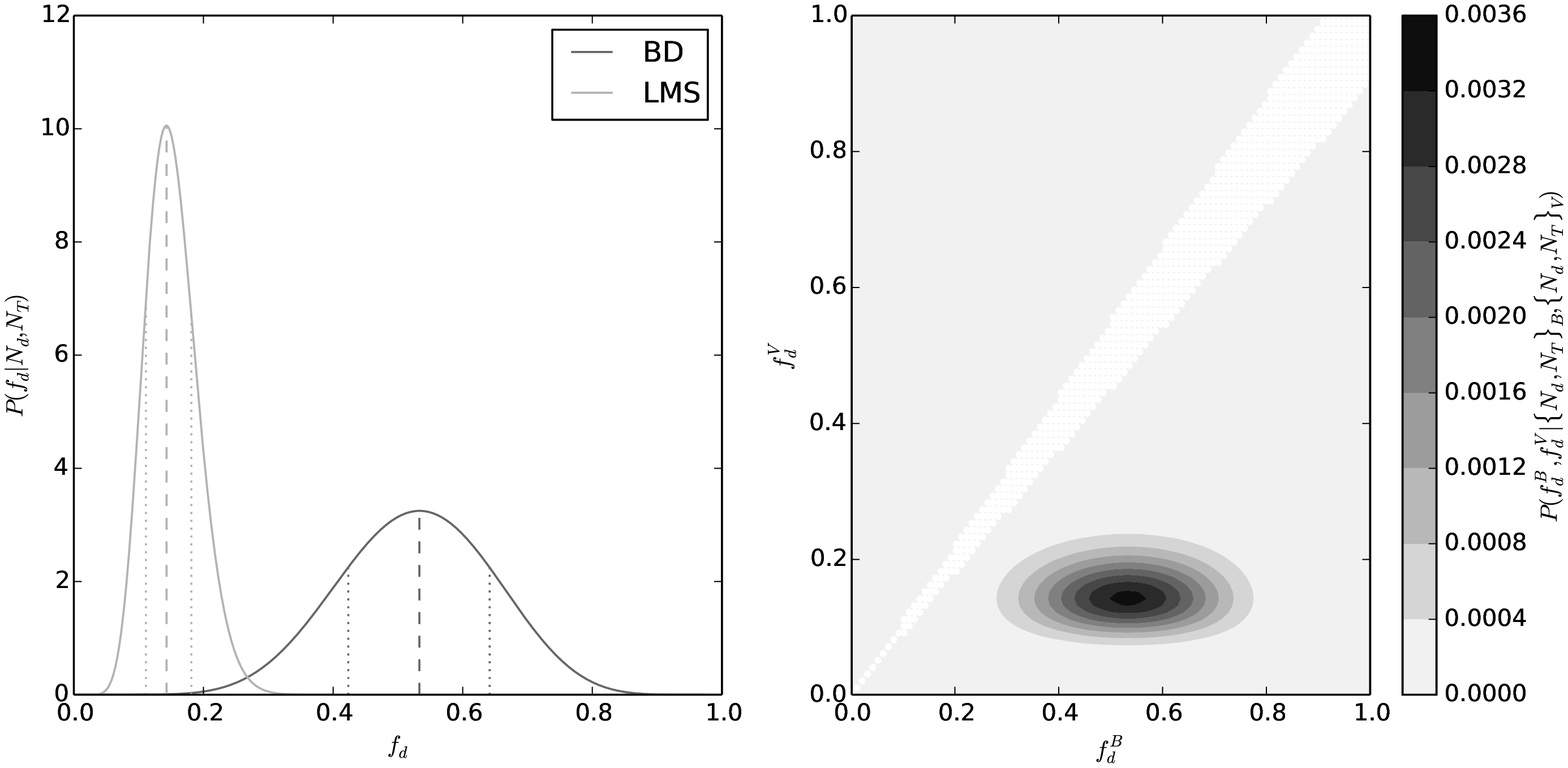}
\caption{Probability distributions for the fractions of members
of class II (upper pannels) and evolved disc (lower pannels).
The left pannels indicate the probability distributions for the fractions of
BDs (solid black curve) and LMS (solid gray curve). The dotted 
vertical lines indicate the $1\sigma$ confidence intervals.
The right pannels indicate the probability distribution (gray scale) as a 
function of the fractions of BD and LMS. See text for the interpretation 
of the probabilities.}
\label{probclassII}
\end{figure}

%
%
Applying this for the LMS and BD samples detailed in the previous section, for class II 
objects we have $N_{d}^{V}=3$, $N_{d}^{B}=3$, $N_{T}^{V}=77$ and $N_{T}^{B}=15$ which results 
in the Posterior PDFs shown in the upper panels of Figure~\ref{probclassII}. The right panel 
shows isocontours of the 2D posterior of Eq. \ref{ec:posterior_vl_bd}. The left panel shows 
the marginal 1D Posterior PDFs for the class II fraction of LMS (gray solid line) and BDs 
(black solid line). The resulting fractions are $f_{d}^{V}=(3.9^{+2.4}_{-1.6})~\rmn{\%}$ and 
$f_{d}^{B}=(20.0^{+9.9}_{-7.9})~\rmn{\%}$, which correspond to the most probable values and 
the respective $1\sigma$ confidence intervals, computed as the 16th and 84th percentiles 
of the corresponding PDFs, respectively shown with dashed and dotted lines in the 
upper left panel of Figure~\ref{probclassII}.
%
%
\par
The next question is what is the probability that these two populations have 
differing class II fractions. It is clear that in this case the obtained disc fractions
are quite different, even taking into account the reported uncertainties, but
the posterior PDF of Eq. \ref{ec:posterior_vl_bd} allows us to quantify this
probability in a general manner, which will be useful for less clear cases.
 Marginalizing the 2D posterior PDF, we compute the probability $P_{>x}$
that the class II fractions of the two populations differ by more than a certain
fractional margin $x$ as:
\begin{equation}\label{ec:Px}
P_{>x}=1-\int_{-x}^{+x} dX\,P(f_{d}^{V},f_{d}^{B})=(1+X)f_{d}^{V}|\{N_{d},N_{T}\}^{V,B})
\end{equation}
In this case, choosing a fractional difference of $0.1$, we find $P_{>0.1}=0.9908$, i.e.
there is a $99.08~\rmn{\%}$ probability that the class II fractions of LMS and BD are
different by more than $10~\rmn{\%}$.
%
%
\par
Similarly, we use Eq. \ref{ec:posterior_vl_bd} to compute the Posterior PDF for the fractions 
of evolved discs and class III objects in the LMS and BD mass regimes. For evolved discs we find 
fractions of $14.3^{+3.8}_{-3.2}~\rmn{\%}$ and $53.3^{+10.9}_{-11.0}~\rmn{\%}$ for LMS and BDs respectively,
with a probability $P_{>0.1}=0.9987$ (from Eq. \ref{ec:Px}) that these two fractions differ by 
more than $10~\rmn{\%}$ (Figure \ref{probclassII}). For class III we obtain fractions of $81.8^{+3.8}_{-4.0}~\rmn{\%}$ and 
$26.7^{+10.4}_{-9.0}~\rmn{\%}$ for LMS and BDs respectively, with a probability $P_{>0.1}=0.9988$. 
The number fractions and $P_{0.1}$ probabilities are summarized in Table \ref{nir}.
%
%
We conclude that the number fractions of class II, evolved discs and class III clearly differ 
at both sides of the sub-stellar mass limit. This differences are statistically robust as 
supported by the $P_{>0.1}$ probabilities. 
%
%
\par
Finally, we emphasize that we confirmed the number fractions found by \citet{downes2014a} 
in the 25 Orionis group based only on photometric BD candidates, with a sample of spectroscopically 
confirmed members. The fractions we obtained following the Bayesian procedure for LMS are also 
consistent with those reported by \citet{briceno2005}, \cite{briceno2007a} and \citet{downes2014a}. 
The final disc classification of each BD is shown in Table \ref{physprop}.
%
%
\section{Spectroscopic signatures of accretion}\label{accretion}
%
%
The evidence of discs around BDs comes not only from IR excesses
but also from signatures related to magnetospheric accretion such as
broad permitted emission line profiles and optical continuum veiling
\citep[e.g.][]{white2003,jayawardhana2003,muzerolle2005,luhman2012}.
The number fraction of BDs classified as CTTS sub-stellar analogous 
is also an indication of the evolutionary stage of its population in the 
sense that this number fraction is expected to decrease with time.
In this section we study the equivalent width of H$\alpha$ as a function 
of the spectral type in order to classify the new BDs as sub-stellar analogous 
of the WTTS or CTTS. Also, we compare these results with those from the IR excesses 
indicative of discs discussed in Section \ref{discs} as well as with stellar 
counterparts.
%
%
\par
\begin{figure}
\includegraphics[width=80mm]{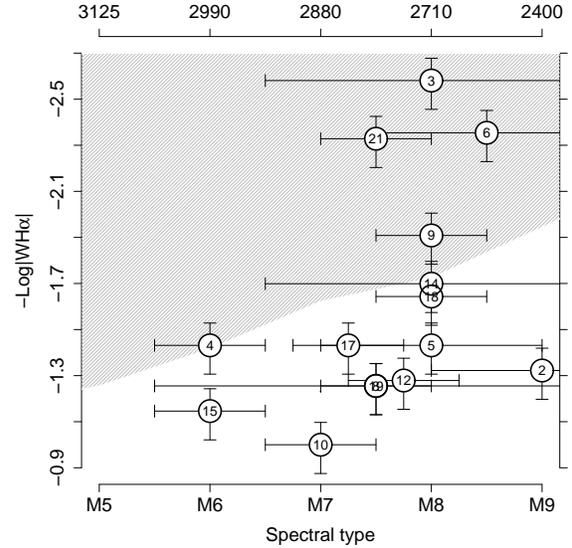}
\caption{Equivalent width of H$\alpha$ versus spectral type for the BDs
confirmed as members. The shadowed area indicates the H$\alpha$ emissions 
above the maximum expected from chromospheric activity \citep{barradoynavascues2003}.}
\label{whast}
\end{figure}

We classified the new BDs as sub-stellar analogous of the WTTS or CTTS
based on the equivalent width of the H$\alpha$ line in emission
and their spectral types, according to the empirical classification scheme
from \citet{barradoynavascues2003} in which a BD is a CTTS analogue
if its H$\alpha$ emission is stronger than is expected from purely chromospheric 
activity which depends on their spectral type (see Figure~\ref{whast}).
%
%
There are several optical spectral features which indicate ongoing accretion in LMS and BDs, 
such as H$\beta$, H$\gamma$, HeI $\lambda$5876 lines in emission as well as the veiling of 
the photospheric absorption lines \citep{muzerolle2003,muzerolle2005}.
%
%
However, the low SNR in the blue-most range of our spectra does not allow for 
a reliable detection of most of these additional lines or to measure veiling. 
The equivalent widths of H$\alpha$ are shown in Table \ref{speccand} 
and the corresponding close up of the H$\alpha$ line for each BD in Figure~\ref{spectrum}.
%
%
Figure~\ref{whast} shows the H$\alpha$ equivalent width as a function of
the spectral type for the new BDs as well as the limit between WTTS and
CTTS proposed by \citet{barradoynavascues2003}. We find that 5 of the new
BDs show signatures of active accretion consistent with CTTS and that the
remaining 10 BDs of the sample show low H$\alpha$ emissions as expected
in WTTS. Following the procedure explained in Section \ref{fractions}, these
corresponds respectively to CTTS and WTTS fractions of $33.3^{+10.8}_{-9.8}~\rmn{\%}$ and
$66.7^{+9.7}_{-10.9}~\rmn{\%}$ among BDs (see Table \ref{nir}).
\begin{table}
\centering
\begin{minipage}{800mm}
\caption{Summary of properties for the new confirmed BDs.}
\begin{tabular}{lccccc}
\hline
ID     & Log($L/L_\odot$) & T$_{eff}$    &  Mass\footnote{Masses were estimated from the interpolation of the\\ $T_{eff}$ 
and luminosities into the DUSTY models from \citet{chabrier2000}.}     & TTS      & Classification \\
       &                  & [K]          &  [$\rmn{M}_\odot$]     &          &               \\
\hline
 2     & -2.7808          & 2400         &       0.01       & WTTS     & Evolved      \\ 
 3     & -2.5400          & 2710         &       0.03       & CTTS     & Class II  \\
 4     & -1.7326          & 2990         &       0.08       & CTTS     & Class II      \\
 5     & -2.6480          & 2710         &       0.03       & WTTS     & Evolved       \\
 6     & -2.6024          & 2577         &       0.02       & CTTS     & Evolved       \\
 8     & -2.4048          & 2808         &       0.04       & WTTS     & Evolved     \\
 9     & -2.6476          & 2710         &       0.03       & CTTS     & Evolved     \\
10     & -2.2248          & 2880         &       0.06       & WTTS     & Class III     \\
12     & -2.0460          & 2808         &       0.03       & WTTS     & Class III     \\
14     & -2.6484          & 2710         &       0.03       & WTTS     & Evolved      \\
15     & -2.1845          & 2990         &       0.08       & WTTS     & Class III   \\
17     & -2.1508          & 2880         &       0.05       & WTTS     & Evolved     \\
18     & -2.4640          & 2710         &       0.03       & WTTS     & Class II      \\
19     & -2.3276          & 2880         &       0.06       & WTTS     & Class III     \\
21     & -2.4896          & 2808         &       0.04       & CTTS     & Evolved      \\
\hline
\end{tabular}
\label{physprop}
\medskip
\end{minipage}
\end{table}
%
%
\par
\begin{figure}
\includegraphics[width=90mm]{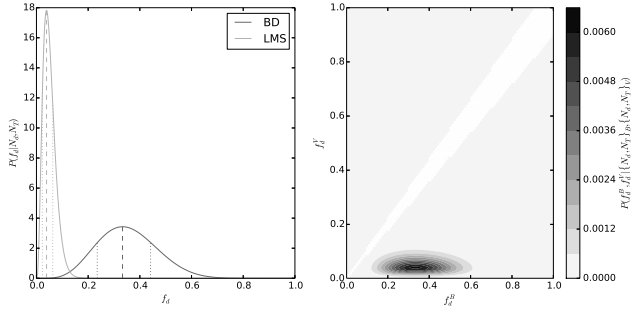}
\caption{Same as Figure~\ref{probclassII} for the fractions of BD and LMS classified as CTTS.}
\label{probctts}
\end{figure}

We classified the 77 LMS from \citet{downes2014a} as CTTS or WTTS following the same procedure 
we used for BDs. We find 3 CTTS (spectral types M1, M1.5 and M5.5) and 74 WTTS (spectral types 
between M0.5 and M5.5) corresponding respectively to $3.9^{+2.4}_{-1.6}~\rmn{\%}$ and
$96.1^{+1.7}_{-2.3}~\rmn{\%}$ which are consistent with the results from 
\citet{briceno2005} and \citet{downes2014a} in the same region.
%
%
As we did for the IR excesses, in Figure~\ref{probctts} we show the probability distribution of the 
fraction of CTTS in the LMS and BD mass regimes. We found that within a $10~\rmn{\%}$ margin there are, 
respectively, probabilities of $0.9993$ and $0.9809$ that the fractions of CTTS and WTTS for BDs have 
different values than for LMS. Again, these probabilities show that the difference observed in the CTTS and 
WTTS fractions obtained for LMS and BDs has a high statistical significance.
%
%
\par
We emphasize that from the 5 BDs showing accretion signatures, 2 were classified as
class II and 3 as evolved discs. In the LMS regime from the 3 objects classified as CTTS, 
2 show IR excesses indicative of class II and 1 was classified as an evolved disc. 
Even if we considered as accretors only those BDs that were classified as CTTS \emph{and} 
class II, the fractions at both sides of the sub-stellar mass limit remain different: 
$13.3^{+8.8}_{-6.4}~\rmn{\%}$ for BDs, $2.6^{+1.9}_{-1.3}~\rmn{\%}$ for LMSs and with a probability 
$P_{0.1}=0.9833$ that both fractions are different within $10~\rmn{\%}$. Finally, the BD 4  
was classified as CTTS although it is close to the WTTS/CTTS limit from 
\citet{barradoynavascues2003}. Even if we considered as CTTS only the BDs showing strong 
$H_\alpha$ in emission (numbers 3, 6, 9 and 21), the CTTS fraction is $26.7^{+10.4}_{-9.0}$ 
which is still higher than the fraction obtained in the LMS regime. 
\section{Comments on particular objects}\label{objects}
$(i)$ The BDs 4, 10, 12, 15, 17 and 19 have information in the J, H and Ks-bands 
from the surveys VISTA and 2MASS. From these, the BDs 19 and 17 show evidence of 1$\sigma$ 
variability in the three band-passes but such level of variability does not 
change the general results presented here because the fits to the \citet{allard2012}
models, changing the photometry randomly within the variations we found, do not
change significantly.

$(ii)$ The BD 4 shows an unexpectedly high extinction $A_V=6.2$, considering that the
mean extinction towards the 25 Orionis group and Orion OB1a is $\bar{A_V}=0.3$. 
This BD shows a SED clearly consistent with the median SED for Taurus and was
classified as a class II although the available photometry at wavelengths shorter 
than H-band, where the excesses are not detected, is not enough for a reliable fit 
into the photospheric models. Additionally it shows $H_\alpha$ line in emission 
which is consistent with ongoing accretion. We speculate that its strong extinction 
could be a result of the disk being edge-on.

$(iii)$ The BD 9 shows an extinction $A_V=1.36$ which is slightly higher but it is 
consistent with mean extinction towards the region if the uncertainty in the 
spectral type is considered.

$(iv)$ The BD 4 was classified as CTTS and the BDs 14 and 18 as WTTS but
there are close to the CTTS/WTTS limit proposed by \citet{barradoynavascues2003}.
New spectroscopic observations with a higher SNR are needed to improve their spectral
classification (particularly BD 14) as well as the measurement of the H$\alpha$ equivalent 
widths in order to improve their classification as CTTS or WTTS sub-stellar analogous.
%
%
%
%
\section{Summary and conclusions}\label{summary}
%
%
We have studied 21 candidate BDs belonging to the 25 Orionis group and the Orion OB1a 
subassociation and spectroscopically confirmed 15 of them as new sub-stellar members 
with spectral types between M6 and M9. Comparing the SEDs of the new members at wavelengths 
beyond $\sim2~\mu\rmn{m}$ with those from the BT-Dusty photospheric models \citep{allard2012} 
for the temperatures corresponding to their spectral types and with the mean SED for Taurus 
representative of class II \citep{furlan2006}, we detect IR excesses indicative of class II, 
evolved disc and class III. We also classified the new BDs as CTTS and WTTS sub-stellar analogous 
according to the empiric limit proposed by \citet{barradoynavascues2003}.
%
%
\par
We have applyed a Bayesian analysis to compute the number fractions of BDs showing 
disc and magnetospheric accretion signatures and compare them with those 
for a sample of 77 LMS belonging to the 25 Orionis group from \citet{downes2014a},
classified following the same procedures.
%
%
We find a significant higher fraction of CTTS, evolved discs and class II objects in 
the sub-stellar mass regime than in the LMS regime  and a fraction of WTTS and class 
III which is lower for BDs than for LMS. All the differences were found to have a high 
statistical significance.
%
%
Our main result is that the number fractions of BDs classified as CTTS \emph{and}
Class II or evolved disc result to be $33.3^{+10.8}_{-9.8}~\rmn{\%}$ for BDs while for the 
LMS is $3.9^{+2.4}_{-1.6}~\rmn{\%}$. Such difference has a probability $P_{<0.1} = 0.9993$ 
of being real.
%
%
\par
The differences between the fractions of discs and accretion signatures at both 
sides of the sub-stellar mass limit could be interpreted, at least, considering 
the following three scenarios: 

$(i)$ If the disc fraction were not dependent on the mass of the central object, 
a way to mimic the observed fractions would be for the formation of BDs to occur 
over a longer period of time than for the LMS. In this scenario, the younger BDs would 
still retain the discs, increasing the mean disc fractions for BDs with respect 
to the LMS. Nevertheless, we found $\sim33~\rmn{\%}$ of BDs are CTTS and class II or evolved 
disc which is similar to the fraction observed for LMS at ages of $\sim3~\rmn{Myr}$ 
\citep[e.g.][$\sigma$ Ori; $\sim35~\rmn{\%}$]{hernandez2007b}. 
Therefore, the observed population should have started to form $\sim7~\rmn{Myr}$ ago and 
extended only for BDs until $\sim4~\rmn{Myr}$ ago. Such an age difference should be reflected 
in the H-R diagram as a larger scatter of the BD locus relative to the LMS, which is not observed 
in the results presented here. In addition, extinction patterns commonly found in 
regions of about $\sim3~\rmn{Myr}$, such as $\sigma$ Ori or Orion OB1b, are not observed 
in the surveyed area \citep{downes2014a}.

$(ii)$ If the time scale for disc dissipation were the same for BDs and LMS, the initial 
fraction of discs would have to be higher for BDs than for LMS, as suggested by 
\citet{riaz2012}, in order to reproduce the larger disc fractions observed for BDs 
relative to LMS at a given age.
Since we are measuring LMS and BD disc fractions at a single age of $\sim7~\rmn{Myr}$,
our observations alone cannot rule out a difference in the initial disc fractions.
However, comparing with observations in the $\sim3~\rmn{Myr}$ old $\sigma$ Ori from 
\citet[][$\sim35~\rmn{\%}$ for LMS]{hernandez2007b} and \citet[][$\sim60~\rmn{\%}$ for BDs]{luhman2008},
we see that for LMS the fraction drops from $\sim35~\rmn{\%}$ to $\sim4~\rmn{\%}$ while for BDs it drops 
from $\sim60~\rmn{\%}$ to $\sim33~\rmn{\%}$. This means the disc fraction drops by a factor $\sim9$ for LMS 
and by a factor of $\sim2$ for BDs in the period of time from $\sim3~\rmn{Myr}$ to $\sim7~\rmn{Myr}$. 
This strongly suggest that, regardless of the initial disc fractions, disc evolution 
occurred faster for LMS than for BDs, which leads us to the third and last scenario.

$(iii)$ The BD and LMS populations are coeval and the time scale for disc evolution depends 
on the mass of the central object, being slower for BDs than for LMS. Our results support this 
scenario, which has been previously suggested by [e.g.] \citet{riaz2012} and \citet{luhman2012b}.
%
%
\par
More spectroscopic and photometric optical and near-IR observations for LMS and BDs populations 
in star forming regions with ages around $\sim10~\rmn{Myr}$ old and older are needed in order to clearly 
constrain how the evolution of the discs proceeds and what is its dependency with the mass of 
the central object. In order to understand the physical scenarios involved, modeling and simulations 
are also needed, as well as a consistent statistical estimation of the number fractions and their 
comparisons for different regions.
%
%
%
%
\section*{Acknowledgments}
J. J. Downes and C. Rom\'an-Z\'u\~niga acknowledges support from Consejo Nacional de 
Ciencia y Tecnolog\'ia de M\'exico (CONACYT) grant number 152160. C. Mateu acknowledges 
support from the postdoctoral Fellowship of DGAPA-UNAM, M\'exico.
\par
We thank the assistance of the personnel, observers, telescope operators
and technical staff at GTC and CIDA, who made possible the observations
at the Gran Telescopio de Canarias and at the J\"urgen Stock telescope 
of the Venezuela National Astronomical Observatory (OAN), especially Antonio 
Cabrera Lavers, Daniel Cardozo, Orlando Contreras, Franco Della Prugna, Freddy 
Moreno, Richard Rojas, Gregore Rojas, Gerardo S\'anchez, Gustavo S\'anchez 
and Ubaldo S\'anchez.
\par
We thank Gladis Magris at CIDA for useful comments that helped improve the 
explanation of the Bayesian technique presented here.
\par
Based on observations made with the Gran Telescopio Canarias (GTC), installed
in the Spanish Observatorio del Roque de los Muchachos of the Instituto de Astrof\'{\i}sica
de Canarias, in the island of La Palma.
\par
Based on observations obtained at the Llano del Hato National Astronomical Observatory of 
Venezuela, operated by Centro de Investigaciones de Astronom{\'i}a (CIDA) for the Ministerio 
de Educaci\'on, Ciencia y Tecnolog{\'\i}a.
\par
This work is based [in part] on observations made with the Spitzer Space Telescope,
which is operated by the Jet Propulsion Laboratory, California Institute of Technology under
a contract with NASA. Support for this work was provided by NASA through an award issued
by JPL/Caltech.
\par 
This publication makes use of data products from the Wide-field Infrared Survey
Explorer, which is a joint project of the University of California, Los Angeles,
and the Jet Propulsion Laboratory/California Institute of Technology,
funded by the National Aeronautics and Space Administration.
\par
Funding for the SDSS and SDSS-II has been provided by the Alfred P. Sloan Foundation,
the Participating Institutions, the National Science Foundation, the U.S. Department
of Energy, the National Aeronautics and Space Administration, the Japanese Monbukagakusho,
the Max Planck Society, and the Higher Education Funding Council for England. The SDSS
Web Site is http://www.sdss.org/. The SDSS is managed by the Astrophysical Research
Consortium for the Participating Institutions. The Participating Institutions are the
American Museum of Natural History, Astrophysical Institute Potsdam, University of Basel,
University of Cambridge, Case Western Reserve University, University of Chicago,
Drexel University, Fermilab, the Institute for Advanced Study, the Japan Participation
Group, Johns Hopkins University, the Joint Institute for Nuclear Astrophysics, the
Kavli Institute for Particle Astrophysics and Cosmology, the Korean Scientist Group,
the Chinese Academy of Sciences (LAMOST), Los Alamos National Laboratory,
the Max-Planck-Institute for Astronomy (MPIA), the Max-Planck-Institute for
Astrophysics (MPA), New Mexico State University, Ohio State University, University
of Pittsburgh, University of Portsmouth, Princeton University, the United States Naval
Observatory, and the University of Washington.
\par
This work makes extensive use of the following tools: TOPCAT and STILTS available
at http://www.starlink.ac.uk/topcat/ and http://www.starlink.ac.uk/stilts/, R from
the R Development Core Team (2011) available at http://www.R-project.org/ and described
in \emph{R: A language and environment for statistical computing} from R Foundation
for Statistical Computing, Vienna, Austria. ISBN 3-900051-07-0, IRAF which is
distributed by the National Optical Astronomy Observatories, which are operated by
the Association of Universities for Research in Astronomy, Inc., under cooperative
agreement with the National Science Foundation and the Virtual Observatory Spectral
Energy Distribution Analyzer (VOSA), developed under the Spanish Virtual Observatory
project supported from the Spanish MICINN through grant AyA2008-02156.
%
%
%

%
%
\bsp
\label{lastpage}
\end{document}